\documentclass[12pt]{book}

\usepackage[dvips]{graphicx,color}
\usepackage{makeidx,tsukuba}

\makeauthorindex
\makeindex

\begin{document}

\BookTitle{\itshape The 28th International Cosmic Ray Conference}
\CopyRight{\copyright 2003 by Universal Academy Press, Inc.}
\pagenumbering{arabic}

\chapter{
Observation of M87 with the Whipple 10m telescope}

\author{%
%
%
S.~LeBohec,$^{1,2}$ I.H.~Bond, P.J.~Boyle, S.M.~Bradbury, J.H.~Buckley,
D.~Carter-Lewis, O.~Celik, W.~Cui, M.~Daniel, M.~D'Vali,
I.de~la~Calle~Perez, C.~Duke, A.~Falcone, D.J.~Fegan, S.J.~Fegan,
J.P.~Finley, L.F.~Fortson, J.~Gaidos, S.~Gammell, K.~Gibbs,
G.H.~Gillanders, J.~Grube, J.~Hall, T.A.~Hall, D.~Hanna, A.M.~Hillas,
J.~Holder, D.~Horan, A.~Jarvis, M.~Jordan, G.E.~Kenny, M.~Kertzman,
D.~Kieda, J.~Kildea, J.~Knapp, K.~Kosack, H.~Krawczynski, F.~Krennrich,
M.J.~Lang, E.~Linton, J.~Lloyd-Evans, A.~Milovanovic,
P.~Moriarty, D.~Muller, T.~Nagai, S.~Nolan, R.A.~Ong, R.~Pallassini,
D.~Petry, B.~Power-Mooney, J.~Quinn, M.~Quinn, K.~Ragan, P.~Rebillot,
P.T.~Reynolds, H.J.~Rose, M.~Schroedter, G.~Sembroski, S.P.~Swordy,
A.~Syson, V.V.~Vassiliev, S.P.~Wakely, G.~Walker, T.C.~Weekes,
J.~Zweerink \\
{\it
(1) Dept. of Physics,Iowa State University, Ames, IA, 50011-3160, USA\\
(2) The VERITAS Collaboration--see S.P.Wakely's paper} ``The VERITAS
Prototype'' {\it from these proceedings for affiliations}
}

\section*{Abstract}
The Whipple 10-m telescope was used to observe M87 since 2000. No significant 
gamma-ray signal was found and upper limits compared to the HEGRA detection 
suggest the source may be variable. We found weak evidence for a 
correlation with the X-ray activity in 2000-2001 but this tendency did 
not persist in 2002-2003.
 
\section{Introduction}
The giant radio galaxy M87 has been 
proposed as a  gamma-ray emitting candidate through various mechanisms. At 
a distance of only 16Mpc, the central region is driving a very extended 
$10^{44}\rm erg~s^{-1}$ jet [12] probably powered by 
accretion onto  
the central super massive $\rm \sim 3\times 10^9 M_\odot$ [9]  
black hole. The high 
energy jet has been considered as a possible source of 
gamma rays. Bai \& Lee [4] predicted a gamma-ray flux 
at a detectable level. 
Very recently, Protheroe et al. 2002, gave a range of gamma-ray flux 
predictions in the context of the Synchrotron Blazar Model. In both cases, 
M87 is considered as a misaligned BL-Lac object. In the case of M87, the 
jet is seen at an angle of $\rm \sim 30^o$ [6] and 
the detection of gamma rays from M87 would open up a new class of AGN for 
TeV studies. M87 is also interesting as a potential source of Ultra High 
Energy Cosmic Rays. It was suggested that most cosmic rays with energies 
larger than $\rm 5\times 10^{19} eV$ could 
be coming from M87 after having been deflected by our galactic wind 
[2]. M87 has also been 
proposed as a target toward which one could search for gamma-rays from 
annihilating super-symmetric dark matter[5].
\section{Observation with the Whipple 10-m telescope}
M87 has been 
recently observed both by the HEGRA collaboration and by the Whipple 10-m 
telescope 
collaboration. Both collaborations originally reported upper limits 
[8,11].
The HEGRA collaboration, after applying a more sensitive analysis method, 
recently reported a $4$ standard deviation ($\sigma$) detection above 
$\rm 730~GeV$ 
[1]. The HEGRA observations (83.4 hours) 
were carried out before 2000. The original Whipple 10-m upper limit came from 
observations in 2000 and 2001. We continued observing M87 in 2002 and 2003. In 
total we have now 29 hours of data accumulated over the last 
four years. The standard analysis with a range of energy thresholds was 
applied to this data set. No significant excess was found and upper limits 
were derived.
\begin{figure}[t]
  \begin{center}
    \includegraphics[height=15pc]{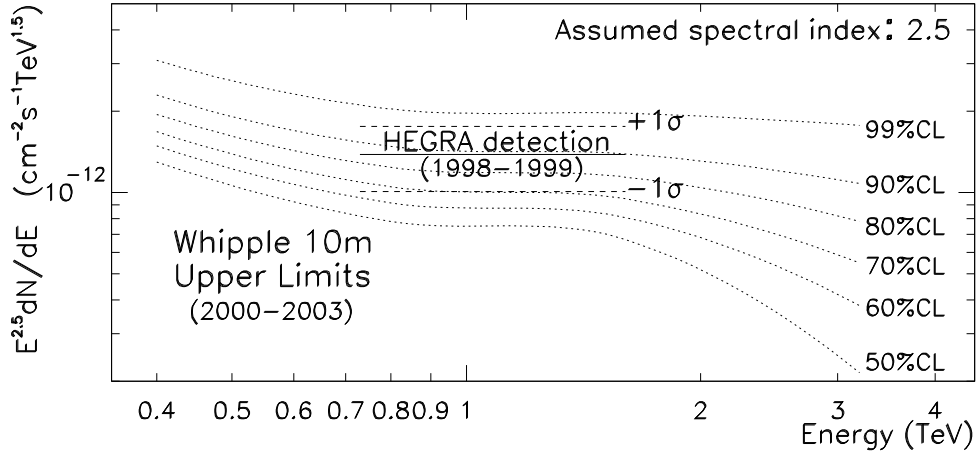}
  \end{center}
  \vspace{-0.5pc}
  \caption{The Whipple 10-m upper limit on the differential flux 
from M87 compared to the detection by HEGRA under the assumption that the 
spectrum ca be described by a power law of index 2.5. }
\end{figure}
This is compared to 
the $4\sigma$ detection by HEGRA on figure 1. The disagreement is marginal, 
but it suggests that the gamma-ray activity of M87 may have reduced since 
the HEGRA observations taken mostly in 1999. We note that the 2000-2001 
data shows a $2.4 \sigma$ excess. 
\section{Correlation between gamma rays and X-rays}
For each 28 minute run taken in the direction of M87 from 2000 until 2003 
we recorded the one day average rate measured with ASM 
(http://xte.mit.edu). The data set was subdivided into 
two  sub-sets corresponding to the periods 2000-2001 (34 runs) and 2002-2003 
(30 runs). On 
figure 2, the gamma-ray rate is shown for each run as a function of 
the X-ray rate for the two subsets. We noticed that some of the points with 
the largest X-ray rates have abnormally large error bars, making the X-ray 
detection low in significance. These points are shown in dashed lines on figure 2 and 
were eliminated from this preliminary analysis as they result from dwell numbers 
of less 
than 5 while corresponding to more than 2 $\rm counts~ min^{-1}$. It must be 
stressed that no 
single run shows a significant gamma-ray excess. Nevertheless 
it seems that the small excess observed in 2000-2001 results mostly 
from the runs obtained on days for which the average ASM X-ray rates were significantly
the highest.  The curve superimposed on figure 2 is an eye-ball fit of a 
quadratic function to guide the eye. The correlation 
coefficient [3] resulting from the 2000-2001 points is 
$0.6\pm 0.2$ while it 
is only $0.0\pm 0.3$ in 2002-2003. 
\begin{figure}[t]
  \begin{center}
    \includegraphics[height=15pc]{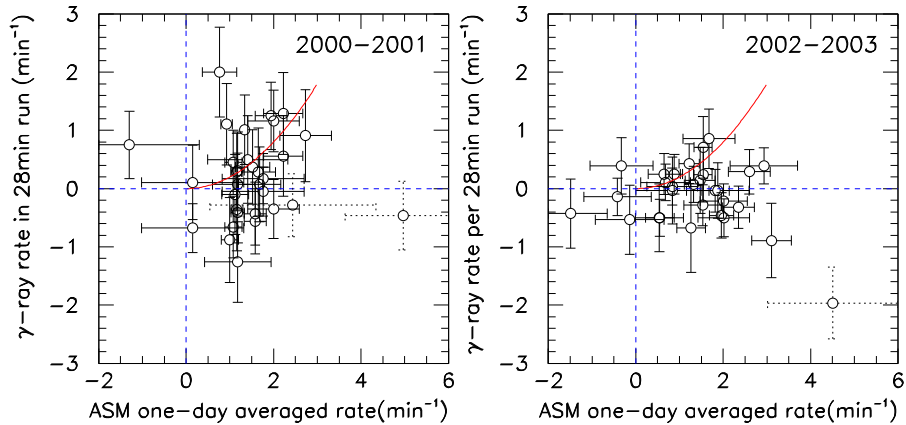}
  \end{center}
  \vspace{-0.5pc}
  \caption{The gamma-ray rate recorded with the Whipple 10-m 
telescope is shown as a function of the X-ray rate provided by the All 
Sky Monitor program in 2000 and 2001 on the left and in 2002 and 2003 on the right. 
 The correlation coefficient in 
2000-2001 is $0.6\pm 0.2$ while it is only $0.0\pm 0.3$ in 2002-2003. 
The curve is an eye ball fit of a quadratic law.}
\end{figure}
The average X-ray rate in the 2000-2001 data set is $\rm 1.24min^{-1}$. We 
constructed the distribution of the $\alpha$ angle for 
runs taken on days on which the X-ray rate was smaller (16 runs) and 
respectively larger (15 runs) than $\rm 1.24min^{-1}$. When M87 was in a low X-ray state we record a $0.6\sigma$ excess while we observe a $3\sigma$  excess 
when the source was in a higher X-ray state. This indication of correlation 
has been a strong motivation for prolonging our observation program in 
2002 but we could not confirm this tendency with the data we recorded 
in 2002 and 2003.
\section{Conclusion}
The 29 hours of data accumulated in the direction of M87 during the last 4 
years do not show any significant gamma-ray excess. The upper limits we 
derived suggest that the gamma-ray activity may have decreased since 
it was
observed and detected by the HEGRA collaboration in 1998 and 1999. A small 
excess ($2.4\sigma$) is noticeable in the data recorded in 2000 and 2001 
with a weak positive correlation with the X-ray activity in M87 as 
measured by ASM. This could not be confirmed in our 
2002-2003 data which does not show any noticeable excess while the 
activity in M87 core and jet was higher [10,13]. 
If the 2000-2001 excess were interpreted as a gamma-ray signal, 
the absence of excess in 2002-2003 could be seen as resulting from an
absorption by the infra-red radiation emitted by 
the heated low temperature torus as suggested by Donea and Protheroe [15].
\section{Acknowledgments}
We acknowledge the technical assistance of E. Roache and J. Melnick. This 
research is supported by grants from the U.S. Department of Energy, by 
Enterprise Ireland and by PPARC in the UK.
\section{References}
\re
1.\ Aharonian et al., 2003, subm. to Astron. and Astrophys., astro-ph/0302155
\re
2.\ Ahn, E., et al., 1999, astro-ph/9911123
\re
3.\ T. Alexander, 1997, in Astronomical Time Series, Eds. D. Maoz, A. Sternberg, and E.M. Leibowitz, (Dordrecht: Kluwer), 163. 
\re
4.\ Bai, J.M., and Lee, M.G., 2001, ApJ, 549, L173.
\re
5.\ Baltz, E.A., et al., 2000, Phys. Rev. D61 023514
\re
6.\ Bicknell,G.V., and Begelman, M.C., 1996, ApJ, 467,597.
\re
7.\ Donea, A.C., and Protheroe, R.J., 2003, astro-ph/0303522
\re
8.\ G\"otting et al., 2001, Proc. of the 27th ICRC, OG 2.3.199, Vol 7, 2669.
\re
9.\ Harms, R.J., et al., 1994, ApJ, 435, L35.
\re
10.\ Harris, D.E., 2003, astro-ph/0302290.
\re
11.\ LeBohec et al., 2001, Proc. of the 27th ICRC, OG 2.3.191, Vol 7, 2643.
\re
12.\ Owen, F.N., et al., 2000, ApJ, 543,611.
\re
13.\ Perlman, E., et al., Central Bureau for Astronomical Telegrams, Circ. No. 8075
\re
14.\ Protheroe, R.J., et al., 2002 subm. to Astropart.Phys., astro-ph/0210249.
\re
15.\ Donea, A.C., and Protheroe, R.J., 2003, astro-ph/0303522
\endofpaper
\end{document}